# Vagus nerve manipulation and microglial plasticity in the prenatal brain

Marc Courchesne[1,2], Colin Wakefield[3], Karen Nygard[1], Patrick Burns[4], Gilles Fecteau[4], Andre Desrochers[4], Mingju Cao[6], Martin G. Frasch[5,6,7,8,9]

[1] Biotron, Western University, London, ON, Canada
[2] Department of Biology, Western University, London, ON, Canada
[3] Drexel University College of Medicine, Philadelphia, PA, USA
[4] Dept. of Clinical Sciences, Faculty of Veterinary Medicine, Université de Montréal, QC, Canada
[5] Dept. of Obstetrics and Gynaecology and Dept. of Neurosciences, Faculty of Medicine, Université de Montréal, Montréal, QC, Canada
[6] CHU Ste-Justine Research Centre, Montréal, QC, Canada
[7] Animal Reproduction Research Centre (CRRA), Faculty of Veterinary Medicine, Université de Montréal, Montréal, QC, Canada;
[8] Center on Human Development and Disability, University of Washington, Seattle, WA, USA
[9] Dept. of Obstetrics and Gynaecology, University of Washington, Seattle, WA, USA

**Corresponding author:**
Martin G. Frasch
Department of Obstetrics and Gynaecology
University of Washington
1959 NE Pacific St
Box 356460
Seattle, WA 98195
Phone: +1-206-543-5892
Fax: +1-206-543-3915
Email: mfrasch@uw.edu

**Abstract**

The efferent and afferent effects of the vagus nerve on the developing brain have remained enigmatic. Here we review the evidence of such effects on microglial plasticity in the sheep model of human fetal development, one of the most recognized and deployed models of human fetal physiology. We show that vagotomy alters microglial phenotype and that this effect is hormetic under conditions of mild systemic inflammation, as may occur antepartum with chorioamnionitis. We present the methodology to assess not only biomarker-based microglial activation (Iba-1), but also the morphometric features of the microglia. Together, these assessments provide a more comprehensive toolbox of glial phenotypical characterizations, especially in the context of investigating the locoregional vagal control of glial function. The presented findings support the earlier discoveries in preclinical and clinical models of adult physiology whereby vagotomy appeared neuroprotective for Parkinson's, explained, at least in part, by the effects on microglia. In addition, we present the approach to measure and the findings on regional cerebral blood flow changes in relation to vagus nerve manipulation. Together, the body of evidence underscores the importance of both the efferent and the afferent vagal pathways, via the vagus nerve, in the programming of microglial phenotype in the developing brain. The significance of these relationships for the development and treatment of early susceptibility to neuroinflammatory and neurodegenerative disorders in later life requires further studies.

**Introduction**

The immune system was long believed to operate independently of the central nervous system (CNS). It was understood to operate autonomously of neural input, driven purely through cytokine response cascades. This dogma has been overturned in recent decades due to growing evidence of neuroimmune interactions, in part also known as the cholinergic anti-inflammatory pathway (CAP).[1] The CAP comprises both afferent and efferent vagal nerve signaling to modulate inflammatory responses largely through the control of cytokine release via cholinergic signaling, both in the periphery as well as in the brain.[2,3] Pre-clinical and clinical evidence has established a role for the neuroimmune interactions in a wide variety of biological processes ranging from infection and autoimmunity to cardiovascular disease, cancer, neurodegeneration, and neuromuscular disease. [1,4–6]

*Neuroimmunology: Quis custodiet ipsos custodes?[1]*

The traditional concept of CAP activity refers to the peripheral (i.e., outside the brain) effects of CAP on inflammatory homeostasis, with certain not yet well identified brain regions thought of as providing information processing, sometimes referred to as a neuroimmunological homunculus.[7] Simply put, as the inflammatory response is generated, cytokine receptors found on afferent vagus nerve fibers are activated, sending signals to the nucleus tractus solitarius (NTS) in the brainstem. CNS recognition triggers an efferent vagal response, delivering acetylcholine to the site of macrophage activation in the periphery.[8] Thus, afferent vagal stimulation by cytokine receptors permits CNS recognition and modulation of innate immune activity.

There is however also evidence for a version of the CAP playing a similar regulatory role in neuroinflammation, i.e., in the brain itself. An example of such central effects can be seen in chorioamnionitis, an inflammation typically originating in the mother and affecting the fetal brain.[2,9] In a broad sense, peripheral inflammatory and central, neuroinflammatory, responses are largely precipitated by the activation of macrophages and microglial cells, the primary immune cell of the CNS, respectively. The activation occurs through pattern-recognition receptors (PRR) and toll-like receptors (TLR) found on the cell surface. Receptor activation then triggers an innate immune response, initiating signaling cascades for nuclear localization factors, namely NF-kB, and subsequent production of pro-inflammatory cytokines such as TNF-α, IL-1ß, and IL-6 or HMGB1.[10]

In the brain, acetylcholine is capable of activating α7nAChR receptors found on the surface of microglia and astrocytes, subsequently inhibiting NF-kB (or HMGB1) nuclear translocation and limiting the generation of pro-inflammatory cytokines.[2,9–11] The presence of a similarly centrally acting neuroinflammatory reflex is suggested by pre-clinical data in rats and fetal sheep.[9,12] In animal models, CNS neuroinflammatory responses can be suppressed when immune cells in the CNS, like microglia, are exposed to α7nAChR receptor agonists and acetylcholinesterase inhibitors triggering the same molecular anti-inflammatory cascade outlined above[1]. In this capacity, the CNS is capable of recognizing and limiting the activation of neuroinflammatory responses through CAP peripherally and likely centrally.[1]

---

[1] Who watches the watchers?

Taken together, we propose that two complementary "brain CAP" systems exist that we subsume under the definition of CAP. The first system represents the peripheral afferent sensing of systemic or organ-specific inflammation which, in addition to the traditional central processing and vagal efferent regulatory outflow, provides the brain itself with information about peripheral inflammation in ways that modulate its local neuroinflammatory milieu via central cholinergic signaling on microglial and astrocytic α7nAChR receptors. The second system is central with regard to its afferent and efferent components and likely distributed with direct locoregional sensing and control of inflammation via glial cells and specific neuronal populations.[13] In this dual role, the watcher is watching itself together with the watched in a dynamic spatiotemporal relationship, if we are to circle back to the subtitle of this section.

For completeness sake, it should be mentioned that the neural central control of CAP involves also cholinergic receptors other than α7nAChR, e.g. m1 and m2 muscarinic receptors.[3] The precise mechanisms of the brain's own sensing and control of neuroinflammation remain to be established, especially in the context of development. We review the significance of this neurodevelopmental perspective for lifetime health trajectories in more detail in Desplats et al.[14]

*Cholinergic anti-inflammatory pathway and microglia*

In the following, we focus on the microglial behavior and leave astrocytes for future work. The impact of CAP on microglial cells is of interest due to their distinct role in the innate immune system. Microglial cells are a unique population of macrophages specific to the CNS. Aside from phagocytic activity, microglial cells are responsible for optimizing brain function and tissue maintenance.[15] Early in embryogenesis, microglial cells migrate to the embryonic brain to play a pivotal role in neurogenesis. Animal modeling has demonstrated that in the prenatal period microglial cells promote neuronal death, fasciculation formation, limit axonal outgrowth, regulate laminar positioning, and promote vascularization and myelination of the neuron. Perinatally, microglial cells largely drive neuronal death and survival. Postnatally, microglial cells promote synapse maturation, remodeling, and pruning as well as myelination maintenance.[16] Microglial cells reach a point of maturation one week postnatally. Throughout development and up until this point of maturation, microglial cells maintain an amoeboid morphology characterized by large, round cell bodies and short, thick branches. This developmentally activated phenotype is associated with the variety of functional roles microglial cells possess during the prenatal period. Of interest, upon immune challenge or postnatally in neurodegenerative disease processes, mature microglial cells morphologically take on an amoeboid structure reflective of that seen during prenatal development.[15] This finding contradicts the traditional view that microglial cells adopt either a pro-inflammatory M1 phenotype or an anti-inflammatory M2 phenotype. As evidenced by morphological and genetic findings, it is now apparent that microglial cells have the capacity for a wide spectrum of activation states. This spectrum of activity, reflected by the amoeboid structure, is of interest as recent data have implicated microglial cells in a multitude of CNS pathologies. Moreso, with the presence of the CAP and the expansion of our capacity for vagal nerve modulation, there is enormous potential to take advantage of the functional diversity intrinsic to microglial cells by modulating CAP activity externally via VNS.

To better understand the interaction between CAP and microglial plasticity, our team employed a sheep model to model human fetal development. Sheep models serve as excellent proxies for human physiology due to similarities in blood gas profiles, hormone responses, organ development, and birthing profiles generally comprised of singletons or twins with weights mirroring those of human infants.[11,17,18] Using a biological model that so closely resembles human development allowed our team to explore the relationship between vagus nerve signaling and microglial activity. More specifically we investigated if and how vagotomy and VNS alter the morphometric phenotype of microglial cells, an aspect sometimes neglected in favor of the more readily accessible average area density metrics of a given immunohistological biomarker, such as Iba-1.

*VNS and cerebral blood flow: is there a link between neuroinflammation and rCBF?*

VNS alters regional cerebral blood flow (rCBF).[19] In adults, Conway et al. found VNS-induced increases in rCBF in the bilateral orbitofrontal cortex, bilateral anterior cingulate cortex, and right superior and medial frontal cortex. Decreases were found in the bilateral temporal cortex and right parietal area. Regions of change were consistent with brain structures associated with depression and the afferent pathways of the vagus nerve. As we mention above subsection, Tracey et al. suggested the existence of a neuroimmunological homunculus which implies that neurovascular activity can be measured in certain brain regions reflecting neuroimmunological stimulation or information processing.[7]

Enigmatically, recent studies also revealed that microglia sense and regulate rCBF.[20] The causal order of the connections between the afferent vagus nerve signaling, rCBF dynamics and microglial plasticity remains to be established. Here, we demonstrate the methodological approach and preliminary findings indicating that vagotomy and VNS modulate rCBF in the developing brain experiencing neuroinflammation.

In summary, this chapter provides an overview of the methodological approach to studying microglial plasticity in the developing brain as a function of vagus nerve activity. Next, we report key results focusing on a vulnerable brain region, the hippocampus. We then conclude with a discussion of the implications of these insights for future research and clinical practice.

**Methodology**

Animal care followed the guidelines of the Canadian Council on Animal Care and the approval by the University of Montreal Council on Animal Care (protocol #10-Rech-1560).

*Anesthesia and surgical procedure*

Briefly, we instrumented 57 pregnant time-dated ewes at 126 days of gestation (dGA, ~0.86 gestation) with arterial, venous and amniotic catheters, fetal precordial ECG and cervical bilateral VNS electrodes; 19 animals received cervical bilateral vagotomy (Vx) during surgery of which 8 animals received efferent VNS electrodes and VNS treatment and 7 animals received afferent VNS electrodes and VNS treatment during the experiment.[18,21,22]

Ovine singleton fetuses of mixed breed were surgically instrumented with sterile technique under general anesthesia (both ewe and fetus). In the case of twin pregnancy, the larger fetus was chosen based on palpating and estimating the intertemporal diameter. The total duration of the procedure was about 2 hours. Antibiotics were administered to the mother intravenously (Trimethoprim sulfadoxine 5 mg/kg) as well as to the fetus intravenously and into the amniotic cavity (ampicillin 250 mg). Amniotic fluid lost during surgery was replaced with warm saline. The catheters exteriorized through the maternal flank were secured to the back of the ewe in a plastic pouch. For the duration of the experiment, the ewe was returned to the metabolic cage, where she could stand, lie and eat ad libitum while we monitored the non-anesthetized fetus without sedating the mother. During postoperative recovery antibiotic administration was continued for 3 days. Arterial blood was sampled for evaluation of the maternal and fetal condition and catheters were flushed with heparinized saline to maintain patency. We reported the detailed approach including Vx and VNS elsewhere.[18,21]

*Experimental protocol*

Postoperatively, all animals were allowed 3 days to recover before starting the experiments. On these 3 days, at 9.00 am 3 mL arterial plasma sample was taken for blood gasses and cytokine analysis. Each experiment commenced at 9.00 am with a 1 h baseline measurement followed by the respective intervention as outlined below (Fig. 1). FHR and arterial blood pressure was monitored continuously (CED, Cambridge, U.K., and NeuroLog, Digitimer, Hertfordshire, U.K). Blood and amniotic fluid samples (3 mL) were taken for arterial blood gasses, lactate, glucose and base excess (in plasma, ABL800Flex, Radiometer) and cytokines (in plasma and amniotic fluid) at the time points 0 (baseline), +1 (*i.e.*, immediately after LPS administration), +3, +6, +12, +24, +48 and +54 h (*i.e.*, before sacrifice at day 3). For the cytokine analysis, plasma was spun at 4°C (4 min, 4000g force, Eppendorf 5804R, Mississauga, ON), decanted and stored at -80°C for subsequent ELISAs. After the +54 hours (Day 3) sampling, the animals were sacrificed. Regional cerebral blood flow (rCBF) was measured at selected time points as outlined in the dedicated subsection. Fetal growth was assessed by body, brain, liver and maternal weights.

Lipopolysaccharide (LPS)-induced inflammation in fetal sheep is a well-established model of the human fetal inflammatory response to sepsis.[23–26]

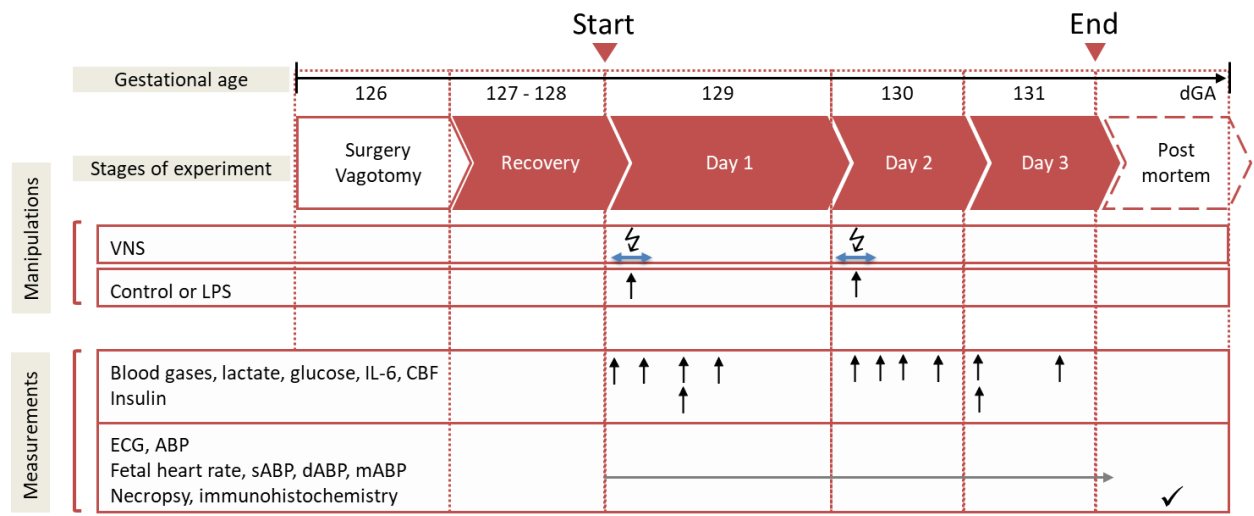

**Figure 1. Experimental design.** Bilateral cervical vagotomy (Vx) was performed during surgery in Vx+LPS group animals. At Days 1 and 2, Vx+LPS animals received LPS dose 400 or 800 ng/fetus/day. Some Vx+LPS animals also received the efferent intermittent (to the periphery) VNS on Days 1 and 2 (Efferent or Afferent VNS groups).

The experimental groups consisted of three following categories.

1) <u>Control and LPS groups:</u> Fifteen fetuses were used as controls receiving NaCl 0.9%. 23 fetuses received LPS (100 n=2, 200 n=1, 400 n=15 or 800 n=5 as ng/fetus/day) derived from E. coli (Sigma L5293, from E. coli O111:B4, readymade solution containing 1 mg/ml of LPS) were administered intravenously to fetuses on days 1 and 2 at 10.00 am to mimic high levels of endotoxin in fetal circulation over several days as it may occur in chorioamnionitis. As we identified that IL-6 response did not depend on the LPS dose in the applied range,[26] these animals were all considered as one LPS group for statistical comparison purposes.
2) <u>Vx+LPS groups:</u> Eleven animals were vagotomized (Vx) and exposed, similar to the LPS group, to LPS400 (n=6) or LPS800 (n=5).
3) <u>VNS: Efferent (n=8) or Afferent (n=7) VNS groups:</u> Fifteen additional Vx animals were subjected to bilateral cervical VNS applied via NeuroLog's NL512/NL800A using pulse sequence pre-programmed in Spike 2 for 10 minutes prior to and 10 min after each injection of LPS. The VNS settings were as follows: DC rectangular 5 V, 100 uA, 2 ms, 1 Hz according to [27]. VENG was recorded at 10,000 Hz.[28]

*Immunofluorescence staining and microscopy*

The microglial marker ionized calcium-binding adapter molecule 1 (Iba-1) expression was quantified. Fetal brains were perfusion-fixed during necropsy and then harvested and immediately fixed in 4% paraformaldehyde (PFA) for 72 hours as reported elsewhere.[29] Samples were then rinsed daily for 3 days using phosphate-buffered saline (PBS), followed by storage in 70% ethanol at 4°C. Brains were divided into thick coronal sections, processed with Leica TP 1020 Automatic Tissue Processor (Leica Instruments, Nussloch, Germany), and embedded in paraffin using the Leica EG 1160 Paraffin Embedding Center (Leica Microsystems, Nussloch, Germany). The Leica RM2145 Rotary Microtome (Leica Microsystems, Nussloch, Germany) was used to cut 5μm sections collected onto Superfrost Plus microscope slides (Fisherbrand, Waltham, MA). Slides were deparaffinized with three 5-minute washes in xylene and rehydrated through a descending ethanol series (100%, 100%, 90%, 90% and 70%) for 2 minutes each followed by deionized water for 5 minutes. Slides were subjected to a heat-induced epitope retrieval process in 10mM sodium citrate buffer (pH 6.0; Sigma-Aldrich, St. Louis, MO) at 90°C in a 2100-Retriever pressure cooker (Electron Microscopy Sciences, Hatfield, PA) for 15 minutes, then rinsed with PBS 3 times for 5 minutes each. Non-specific protein binding was blocked with Background Sniper protein blocker (Biocare Medical, Pacheco, CA) for 10 minutes. Following a brief PBS rinse, the slides were incubated with the polyclonal rabbit anti-Iba-1 (1:1500 Dilution; Wako Pure Chemical Industries Ltd., Osaka, Japan; #019-19741) diluted in Dako diluent solution (Agilent, Santa Clara, CA) overnight at 4°C in a humidity chamber. After another 3 x 5-minute PBS rinse, slides were incubated for 40 minutes at room temperature with Invitrogen Alexa Fluor 647 goat anti-rabbit IgG (Thermo Fisher Scientific, Waltham, MA). 4′,6-diamidino-2-phenylindole (DAPI,1:300 Dilution in PBS; Thermo-Fisher Scientific, Mississauga, ON) was applied as a counterstain for 2 minutes to visualize cell nuclei. Finally, slides were mounted with Prolong Gold antifade mounting medium (Thermo-Fisher Scientific, Waltham, MA). Two negative controls were performed by replacing the primary antibody step with Dako diluent alone and also with Dako purified pre-immune mouse IgG to rule out non-specific binding.

Images were captured on a Zeiss AxioImager Z1 (Carl Zeiss Canada, North York, Ontario) with the user blinded to the treatment conditions. Using the Zeiss Zen stitch and tile and z-stack functions, whole-brain regions were delineated at low magnification and then imaged in their entirety with a 40x oil immersion lens, through a thickness of 3.96 um, which yielded 7 layers through the section volume at optimal spacing. The regions of interest analyzed for this study were the CA1, CA2/3, and dentate gyrus (DG) of the hippocampus. Eight subset images of 300 x 300 μm were then isolated from each brain region and exported as 7-layer tiff images for analysis. This allowed us to include as full a range as possible of the processes associated with each cell body.

These analyses were performed in two batches, separated in time. We account for this in the quantification of the Iba-1 signal in Results section (cf. "Batch effects" subsection).

*Quantification of Iba-1 signal*

All analysis was performed using Image Pro Premier version 9.3 software (Media Cybernetics, Rockland MD). First, the 7-layer 300x300 µm tiff images of (individually) both the multichannel RGB image and the greyscale Iba-1 channel alone were processed through the extended depth of focus (EDF) using the "large edges" algorithm, to yield single-plane images. This created images in which the processes associated with each cell body could be maximally measured in one single focal plane to help differentiate ramified from activated morphology. To allow the normalization of data to the actual tissue area in each field of view, any black, non-tissue area in the multichannel EDF image was first measured, and then subtracted from the total image area to yield the true total tissue area.

Subsequently, 2 copies of the EDF Iba-1 channel image were made. The first copy was background-corrected to eliminate lighting artifacts typical in widefield imaging that could affect cell recognition, while the second maintained its original intensity data. Using the OTSU minimum variance algorithm, the Iba-1 signal of the "whole-cell" (total of cell bodies cell plus processes) was counted on the corrected images, followed by a repeat count using size and roundness filtering to select only the cell bodies. A 2 pixel object growth at the perimeter was employed as well as a "morpho close" filter with a radius of 5 pixels to encourage the joining of adjacent processes to their cell body of origin, to improve morphological and fractal data. This made measurements more accurate, as the intensity difference tended to initially separate them. The outlines of these whole-cell and cell body counts were stored as geographical features, and re-applied in turn to the unaltered Iba-1 channel EDF images to yield the final data so that intensity data would not be affected by the background correction.

For each hippocampal region, the mean object intensity values from either 'whole cells' or 'cell bodies only' of all 8 subset images were averaged. To derive the mean object intensity for only microglial processes, the difference in total sum intensity was divided by the difference in object counts of whole cells and cell bodies (Figure 2E). Similarly, cell counts derived from cell-body filtering from all 8 subset images were summed and divided by the sum of 'true total tissue area' to derive microglia density per region expressed as cells per $mm^2$. Average microglial soma size was determined by dividing the total 'cell-body' area (in $\mu m^2$) by the cell body count of all 8 subsets per region. Finally, the soma to whole cell ratios (%) were determined using respective sizes in $\mu m^2$ and used as a measurement of microglial activation as previously described by Hovens *et al.*[30]

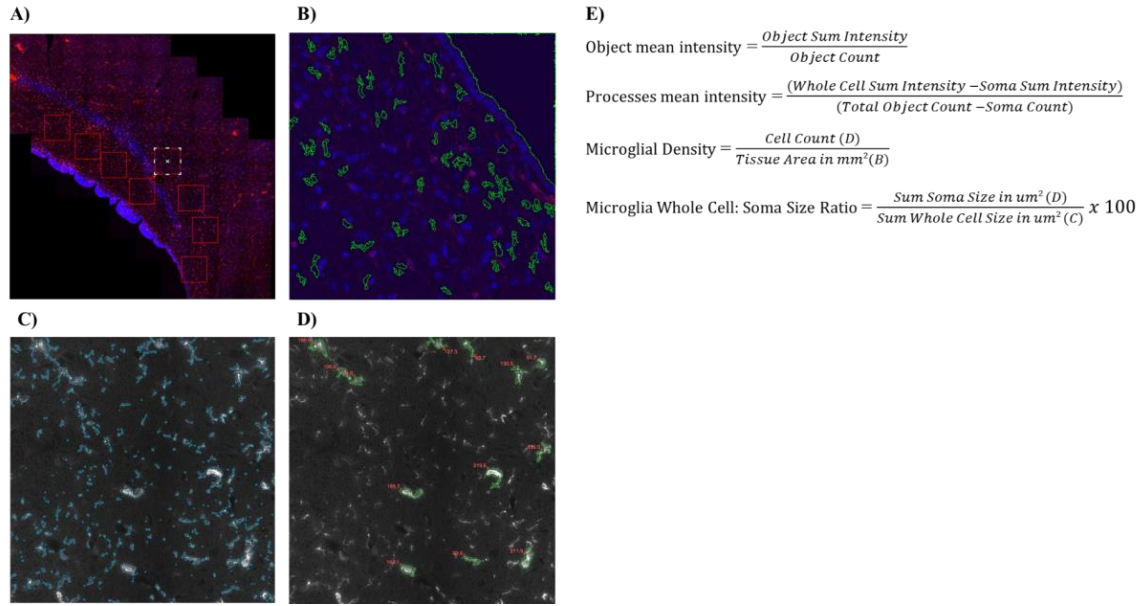

**Figure 2.** Methodology for image analysis and quantification of the immunofluorescent signal. A) Obtaining 8 subsets of images from a single brain region. Red outlines indicate individual 300x300 μm images used for further analysis. B) Removing empty spaces in tissue based on darkness threshold to obtain total tissue area. Green outlines indicate a non-tissue area excluded from further analysis. C) Brightness threshold set to detect whole microglia (Soma and all processes shown with blue outlines. D) Further thresholding (green outlines) to retain cell soma only. E) Formulas used for quantification of Iba-1 object mean intensity, microglia density, and soma: whole cell size ratio.

*Measurement of cerebral blood flow*

Briefly, rCBF was measured at four time points 0 (baseline), +6 (*i.e.*, 6 h after $1^{st}$ LPS), + 24 (*i.e.*, 24 h after $1^{st}$ LPS, just prior to $2^{nd}$ LPS) and +48 h (*i.e.*, 24 h after $2^{nd}$ LPS) with fluorescent microspheres of different colors (NuFLOW Hydro-Coat Microspheres, size 15.5 μm; the concentrations were 5 million/2ml/vial, Interactive Medical Technologies (IMT), CA). To measure rCBF, we used the established fluorescent-colored microsphere (CMS) technique.[31,32] For the purposes of this work, all cortical and subcortical regions were analyzed together.

In the following, we provide a detailed description of the CMS methodology.

To measure CBF, we adopted an established CMS technique using the reference sample method.[33] Microspheres were suspended by gently rolling or swirling the vial, followed by a brief sonication (30 s) with a low-power ultrasound probe, and drawn up into a sterile syringe immediately before injection. The number of injected colored microspheres was estimated to be sufficient to ensure an adequate number of microspheres per tissue sample (> 400) to meet the requirement of a systematic error of about 10 % at the 95[th] confidence level.[34] The minimum number of microspheres per injection was calculated by the following formula:

$CMS_{inj} = CMS_{min} * W_{organ} / [Q_{organ}/Q_{total}]$

Where $CMS_{inj}$ is the microsphere number per injection; $CMS_{min}$ is the minimum number of microspheres required per sample; the value is 400 in the formula.[34] $W_{organ}$ is the average weight of the target organ. In this study, the average fetal brain weight was 52g (range 45-60g); we chose the highest brain weight of 60g for calculation. $Q_{organ}/Q_{total}$ is the percentage of blood flow distributed to the target organ, the cardiac output to the brain in fetal lamb is 15%.[35] So the $CMS_{inj} = 400 \times 60/15\% = 160000$, which is the minimum CMS number required per injection; we optimized practically that an increase to 750,000 per injection would obtain the adequate microspheres count in the current study. Therefore, the volume of each injection (μl) was computed to be 750,000÷(5,000,000/2000μl) = 300μl.

Beginning 15–25 s before the injection of the CMS, a reference blood sample of 4 ml was withdrawn over 2 min from the ascending aorta into a heparinized glass syringe at a rate of 2 ml/min with a syringe pump (Harvard APP, 11 plus 70-2211, single syringe pump). The amount of blood withdrawn for each administration of the CMS for the reference flows (4 ml) was replaced by maternal venous blood. Reference blood samples were stored in the fridge for analysis.

Following the completion of the animal experiment, the right brain hemisphere tissues were processed by brain region, 0.5-2g (average 1g) fetal brain tissue samples were taken from frontal, parietal, occipital, and temporal cortices, white matter, brainstem, cerebellum, and thalamus. Brain tissues and 5 reference blood samples were sent to Interactive Medical Technologies (IMT, CA) for fluorescence analyses. For the purposes of this work, all cortical and subcortical regions were analyzed together, and data was presented in one report for each animal. Then, group averages were calculated.

The rCBF was calculated as follows:

*rCBF (mL/min/gram) = (Total Tissue Spheres) / ((Tissue Weight, Grams) * (Reference Spheres / mL / min))*

*Statistical approach*

Statistical analysis of histomorphometric data was performed using Graphpad Prism 9.4.1 (GraphPad Software Inc, San Diego, CA). Differences in experimental groups were compared in pairs by performing unpaired nonparametric T-tests. The Mann-Whitney test was used to calculate two-tailed p values and differences were considered statistically significant if p<0.05. No adjustments were made for multiple comparisons as recommended by Rothman.[36]

Vertical box and whisker plots depict the median, $25^{th}$, $75^{th}$, and minimum/maximum values. Outliers were determined using Tukey's method and plotted as individual dots. These values are greater than 1.5 times the interquartile range (IQR) from the $75^{th}$ and $25^{th}$ percentile.

For cytokine analyses, general linear modeling (GLM) in Exploratory/R was used to assess the effects of treatment (LPS, Vx+LPS400, Vx+LPS800, Vx+efferent VNS+LPS, Vx+afferent VNS+LPS) while accounting for repeated measurements of fetal plasma IL-6 cytokine. Consequently, experimental groups and time points served as predictor variables; as base levels served the control group and the baseline time point, respectively. For cytokines, the results are presented as median±IQR.

Generalized estimating equations (GEE) modeling was used to assess the effects of LPS while accounting for repeated measurements on fetal rCBF. We used a linear scale response model with time and LPS as predicting factors to evaluate their interactions using maximum likelihood estimate and Type III analysis with Wald Chi-square statistics. SPSS Version 21 was used for these analyses (IBM SPSS Statistics, IBM Corporation, Armonk, NY).

Not all measurements were obtained in each animal. In such a case, the sample size is reported explicitly.

## Results

*Systemic inflammatory response: IL-6*

These findings were reported, in part, elsewhere, except for the effect of the afferent VNS intervention in the Vx+LPS400 treated animals.[22] We provide them here for the proper context in which the brain-specific inflammatory response needs to be interpreted.

LPS provoked a systemic inflammatory response with time effect, i.e., peaking at 3 and 6 hours, as measured by fetal arterial IL-6 concentrations over 3 days (Fig. 1). Surprisingly, Vx+LPS400 restored the levels of IL-6 to control levels, while treatment effects with elevated IL-6 were observed for afferent (but not efferent) VNS, Vx+LPS800, and LPS groups, in the decreasing order of magnitude.

Of note, in the efferent VNS group, the inflammatory response abated quicker than in other LPS-exposed groups, while in the Vx+LPS800 group and even more so in the afferent VNS group, the response persisted the longest, up until 54 hours post LPS, the last experimental measurement time point.

Last but not least, we observed a clearly diametrical temporal profile of the fetal inflammatory response to LPS in Vx+LPS400 versus Vx+LPS800 group, suggesting a sigmoid functional relationship between the vagal cervical denervation and the LPS dose-dependent magnitude of the fetal inflammatory response. We refer to this phenomenon as a hormetic response.[22]

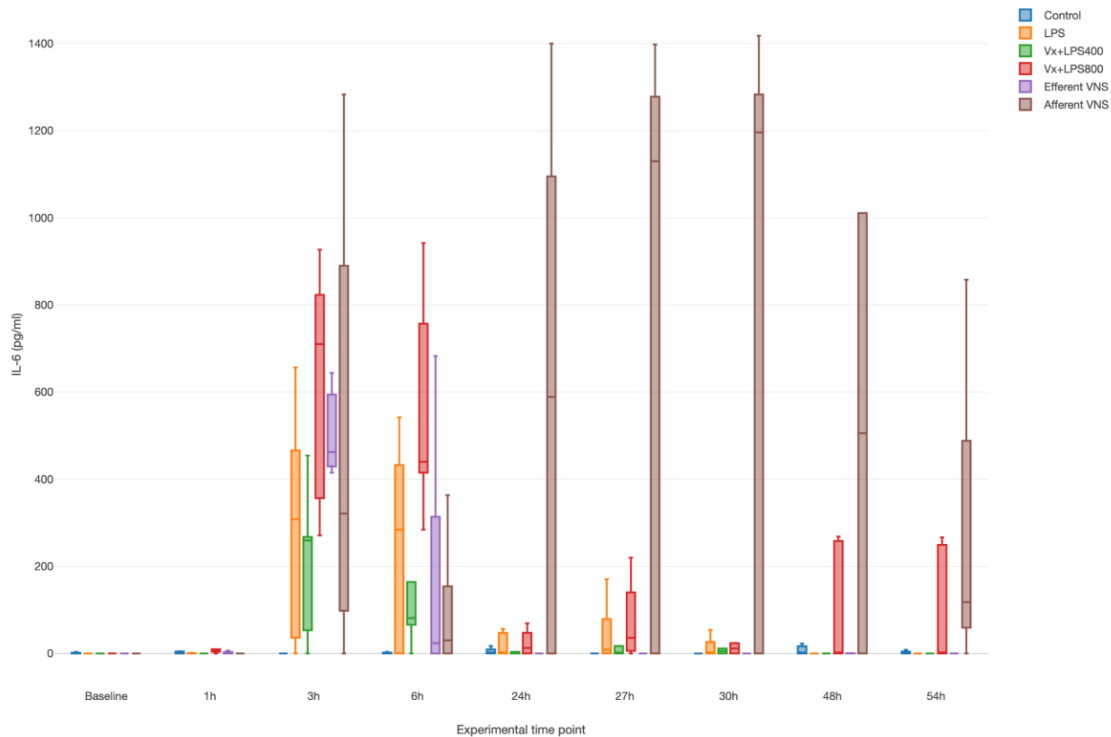

**Figure 3. Fetal systemic inflammatory response to intravenous LPS injection after baseline and at 24 hours: the impact of vagus nerve manipulation.** Vx, bilateral cervical vagotomy during surgical instrumentation; LPS400 and LPS800 indicate the respective intravenous dose of LPS in ng/fetus/day given after baseline and 24 hours later; Efferent and afferent VNS groups received Vx and LPS400 as the Vx+LPS400 group followed by respective targeted VNS treatment around the LPS administration at days 1 and 2. Based on the figure from [22] (reproduced with permission).

*Vagus nerve signaling modulates microglial phenotype*

*Effects of vagus nerve manipulation on Iba-1 expression in the hippocampus*

In this subsection, we focus on the object mean intensity of Iba-1 fluorescent signal in hippocampus subregions. The analyses of the experimental groups from 2 different batches of animals were separated to account for possible intensity differences. We present the analysis of the possible batch effects in the following subsection.

Quantification of fluorescent mean intensity of Iba-1-stained microglia was performed for the whole cell, soma, and processes in the CA1, CA2/3, and DG regions. Signal intensity was decreased for Vx+LPS400 treatment compared to controls ($p=0.04$) for whole microglia only in the CA1. However, once microglial processes were isolated, Iba-1 signal was decreased for Vx+LPS400 relative to controls in the CA2/3 ($p=0.05$) and DG ($p=0.02$). Interestingly, Vx+LPS400 treatment also showed decreased Iba-1 signal intensity compared to LPS only for microglial processes in the CA2/3 ($p=0.04$).

A hormetic effect was observed for vagotomised animals receiving LPS doses of 400 versus 800ng/fetus/day. Iba-1 intensity was increased in Vx+LPS800 compared to Vx+LPS400 in whole microglia of the CA1 ($p=0.04$), CA2/3 ($p=0.03$), and DG ($p=0.02$). For microglial soma only, there was a similar behavior whereby Iba-1 signal in Vx+LPS800 was greater than Vx+LPS400 in the CA1 ($p=0.04$), CA2/3 ($p=0.03$), and DG ($p=0.02$). Finally, when microglial processes were isolated, signal for Vx+LPS800 was increased compared to Vx+LPS400 in the CA1 ($p=0.04$), CA2/3 ($p=0.03$), and DG ($p=0.03$). Notably, no differences of Iba-1 signal mean intensity were observed between controls, efferent and afferent VNS within the same cohort.

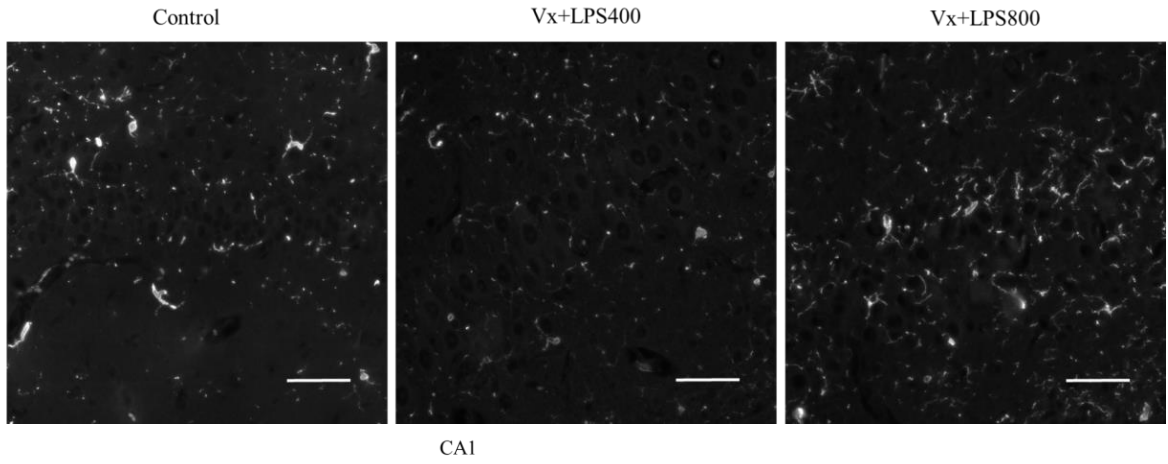

CA1

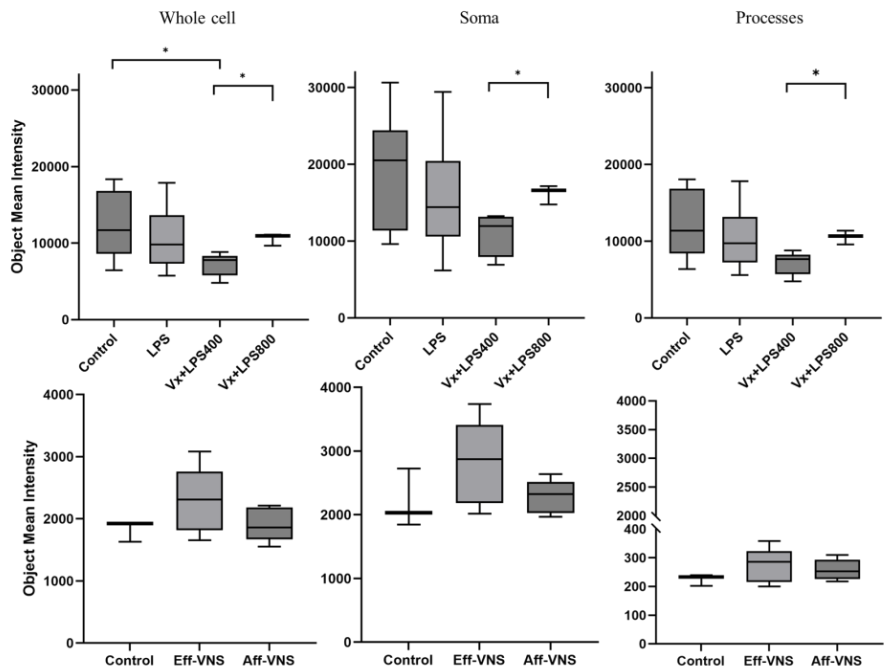

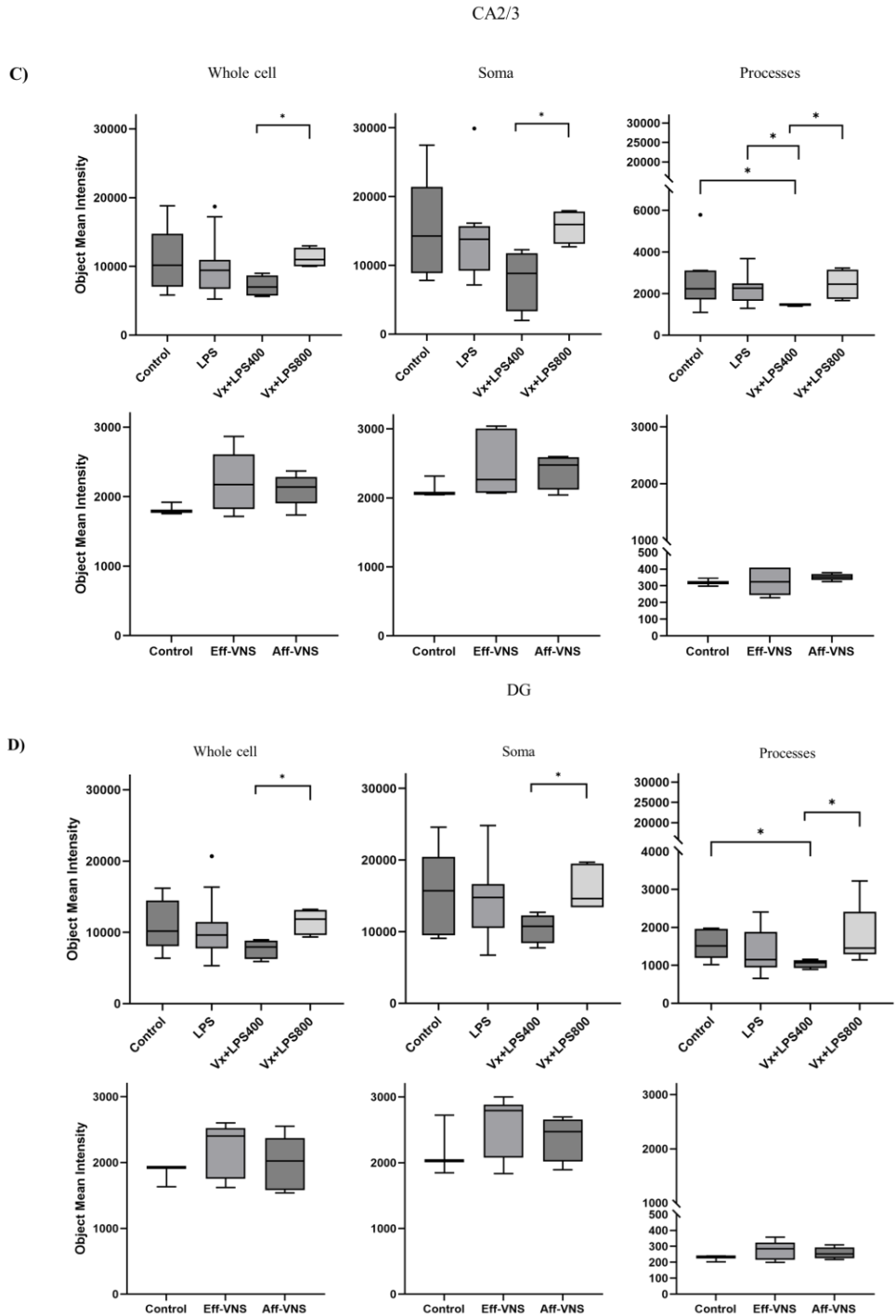

**Figure 4. Effects of systemic vagus nerve manipulation on Iba-1 expression in the hippocampus of near-term fetal sheep. A)** Immunofluorescent staining of Iba-1 on the hippocampal brain tissue of control and Vx+LPS400/Vx+LPS800-treated animals. Representative images of the CA1, CA2/3, and DG sub-regions are shown. Scale bar = 50μm. Object mean intensity of Iba-1 fluorescent signal in the **B)** CA1 (Control n=11, LPS n=13,

Vx+LPS400 n=5, Vx+LPS800 n=3; Control n=3, Eff-VNS n=7, Aff-VNS n=7), **C)** CA2/3 (Control n=8, LPS n=12, Vx+LPS400 n=4, Vx+LPS800 n=4; Control n=3, Eff-VNS n=7, Aff-VNS n=6), and D) DG (Control n=11, LPS n=14, Vx+LPS400 n=4, Vx+LPS800 n=5; Control n=3, Eff-VNS n=7, Aff-VNS n=7). *$p<0.05$.

*Effects of vagus nerve manipulation on microglial morphometry in the hippocampus*

In this subsection, we focus on microglia morphometry data reflecting the cell activation state. Data presented includes cell count per tissue area, average soma size, and percentage of the whole cell occupied by soma only. Groups from 2 different batches of animals were combined. We present the analysis of the possible batch effects in the following subsection.

The microglial soma to whole cell ratio was used as a measure of microglial activation, expressed as a percentage of cell body area relative to the whole cell. A decreased ratio was observed with Vx+LPS400 treated animals compared to controls in the CA1 ($p=0.01$) and DG ($p=0.02$). In the CA2 subregion, Vx+LPS400 treatment was decreased compared to control ($p=0.01$), LPS ($p=0.01$), efferent ($p=0.04$) and afferent ($p=0.01$) VNS.

There was an increase in microglial density following efferent VNS compared to Vx+LPS400 ($p=0.03$) only in the CA1, while no differences between groups were observed in the CA2/3 and DG.

In the CA1, both efferent and afferent VNS lead to a decrease in microglia soma size compared to Vx+LPS400 ($p=0.02$, $p=0.05$), Vx+LPS800 (both $p=0.02$), and LPS only treatments ($p=0.003$, $p=0.02$). Additionally, soma sizes for efferent VNS animals were smaller compared to control ($p=0.03$). In the CA2, Vx+LPS800 treated animals had larger microglia soma size compared to efferent ($p=0.006$) and afferent ($p=0.01$) VNS, while efferent VNS also decreased soma size relative to LPS ($p=0.02$). Similarly, both efferent and afferent VNS lead to smaller soma size relative to LPS animals ($p=0.03$, $p=0.05$) in the DG. Efferent VNS treatment also decreased soma size relative to Vx+LPS800 ($p=0.01$). We did not observe a difference in effects of efferent versus afferent VNS.

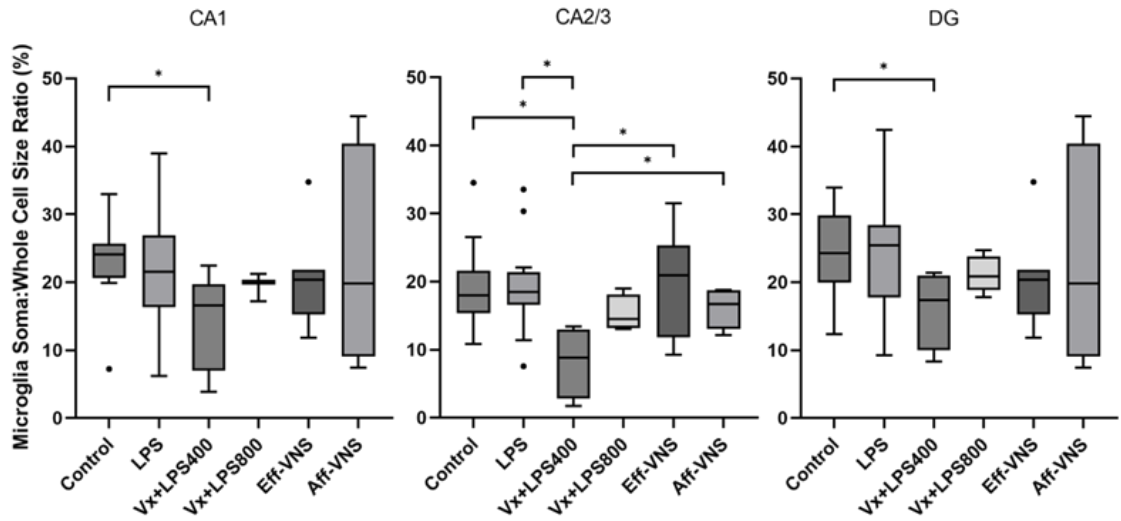
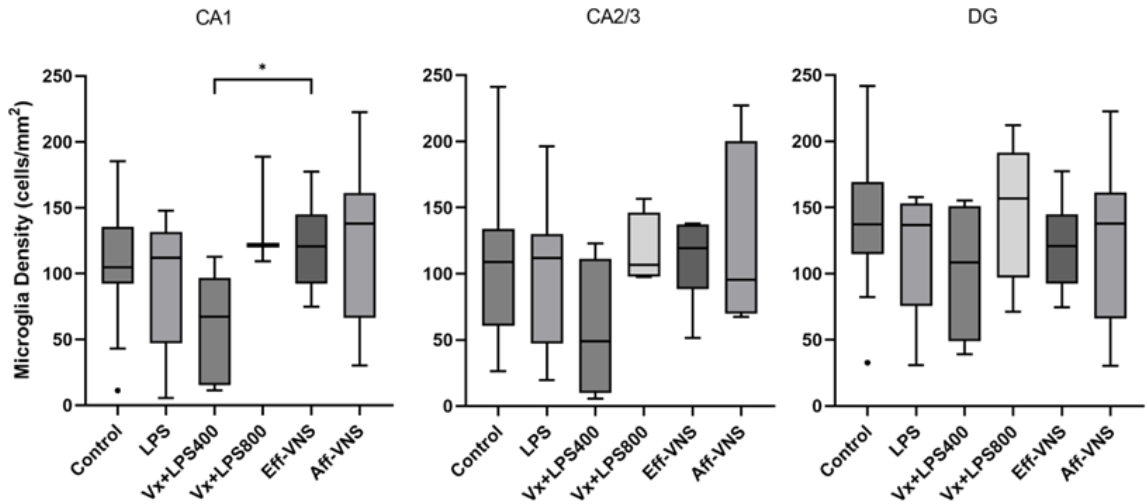
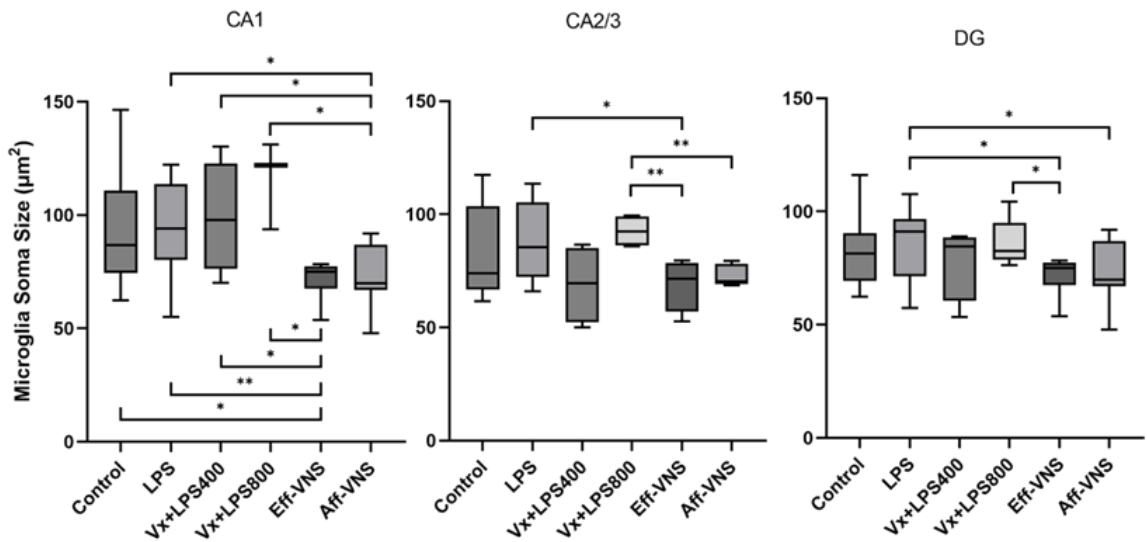

**Figure 5. Effects of systemic vagus nerve manipulation on microglial morphometry in the hippocampus of near-term fetal sheep.** A) Microglia soma to whole cell size ratio represented as a percentage of the whole microglia occupied by cell body only. B) Microglia density expressed as cell body count per mm2 of tissue within the CA1, CA2/3, and DG C) Average microglial soma size in µm2. CA1 (Control n=14, LPS n=13, Vx+LPS400 n=5, Vx+LPS800 n=3, Eff-VNS n=7, Aff-VNS n=7), CA2/3 (Control n=11, LPS n=12, Vx+LPS400 n=4, Vx+LPS800 n=4; Eff-VNS n=7, Aff-VNS n=6), DG (Control n=14, LPS n=14, Vx+LPS400 n=4, Vx+LPS800 n=5, Eff-VNS n=7, Aff-VNS n=7).*p<0.05.

*Batch effects: accounting for measurements of microglial activity made during different time points*

There was a concern about possible signal intensity differences between images captured in 2014 (batch 1) versus 2020 (batch 2), therefore potentially affecting soma size comparisons. However, there were no significant differences with regard to soma size between groups of the same treatment (control & Vx+LPS400). This consistency implies that a fair comparison of soma size can be made between these groups.

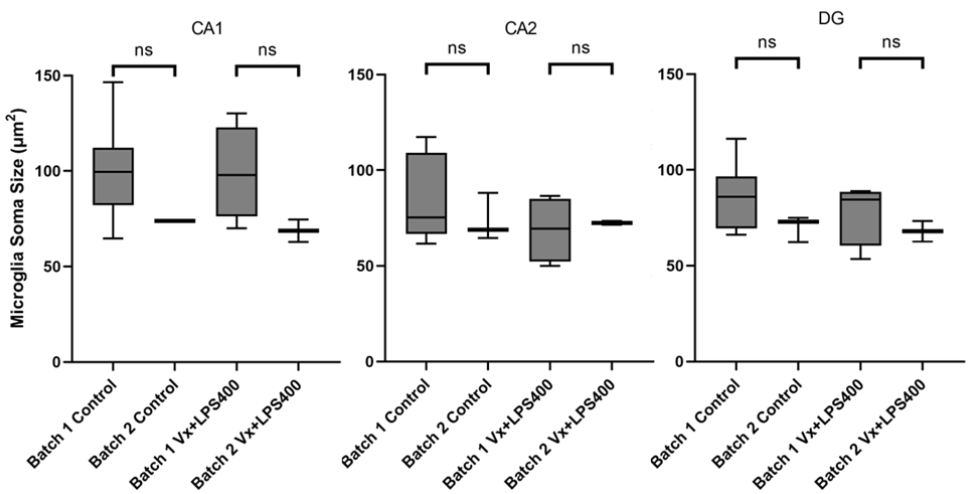

**Figure 6. No differences in microglial soma size were observed between batches.** CA1 (Batch 1 control n=11, Batch 2 control n=2; Batch 1 Vx+LPS400 n=5, Batch 2 Vx+LPS400 n=2), CA2/3 (Batch 1 control n=8, Batch 2 control n=3; Batch 1 Vx+LPS400 n=4, Batch 2 Vx+LPS400 n=2), DG (Batch 1 control n=11, Batch 2 control n=3; Batch 1 Vx+LPS400 n=4, Batch 2 Vx+LPS400 n=2). n.s., p>0.05.

In summary, the key finding is that vagotomy influences microglial plasticity in an LPS dose-dependent hormetic manner.

Consistent with the earlier observations in systemic and organ-specific, regional effects in the terminal ileum,[22] vagotomy exerted a hormetic effect on the fetal brain's microglial cell soma/whole body ratios. Interestingly, this effect was not apparent in whole cell density observations. Consistent with systemic reduction of IL-6 cytokine release, here we observed a reduction in soma/cell size ratio indicative of a less reactive microglial phenotype under conditions of Vx+LPS400, but not Vx+LPS800, below LPS or even control levels. This finding was consistent across all 3 hippocampus regions. In CA2/3, additional effects of efferent and afferent VNS were also observed as compared to Vx+LPS400.

We did not observe a difference in the effects of efferent versus afferent VNS. Both VNS regimes appeared to restore the microglial phenotype in these brain regions to control levels.

*Vagus nerve activity modulates rCBF*

LPS resulted in a delayed (versus the peak of the inflammatory response) onset of hyperperfusion 2x compared to controls: this was seen at 24h in the subcortical matter compared to cortical matter. Similar effects of intracerebral redistribution of blood flow have been observed in this species and gestational age due to umbilical cord occlusions that cause fetal neuroinflammation.[2,37]

Vx+LPS increased this effect 3-fold and appeared as early as 6 hours. Efferent VNS reduced the rCBF to below control levels for the cortical matter, while afferent VNS diminished the rise of rCBF in the subcortical, but not cortical matter.

The above-described changes in rCBF can be quantified as redistribution - a measure of redirected CBF from cortical to subcortical regions. We observed main effects for the treatment groups LPS, Vx + LPS400, Vx + LPS800 (i.e., Vx+LPS as shown in Figure 6), and Vx+Efferent VNS+LPS ($p=0.000$, $p=0.03$, $p=0.045$, and $p=0.02$ respectively). In addition, we also found main effects for the time points "24 hours" and "48 hours" ($p=0.000$ for each).

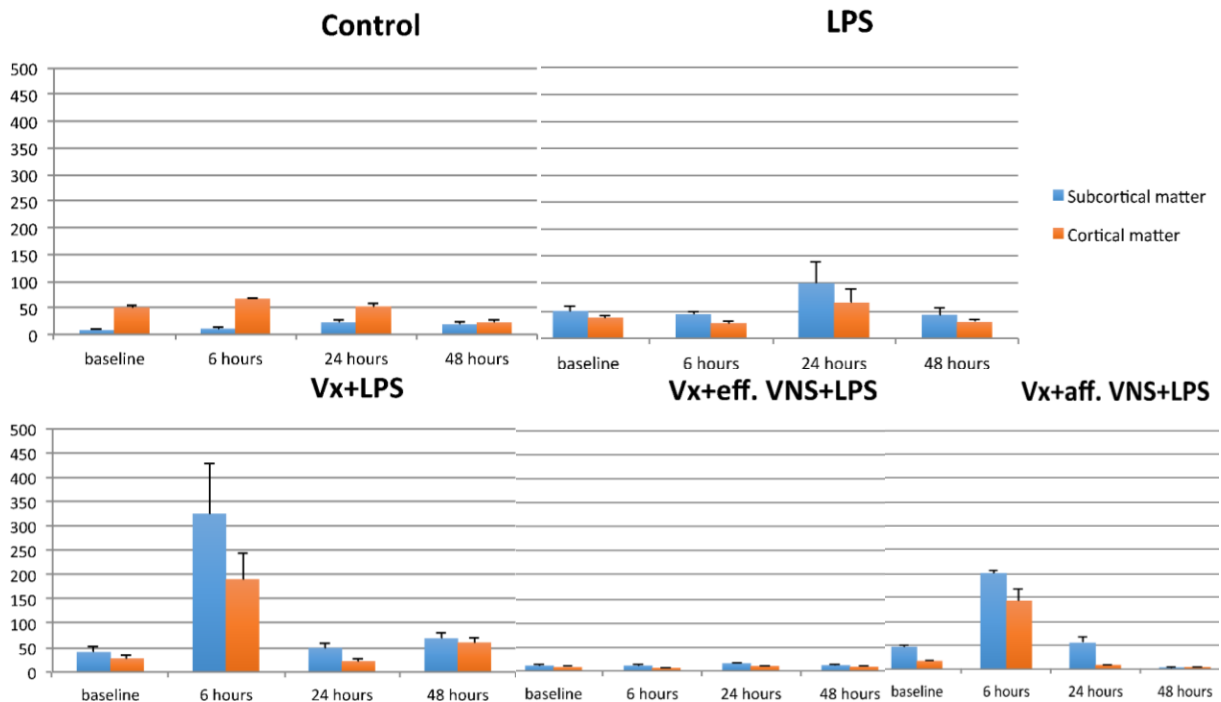

**Figure 7. Regional cerebral blood flow** in cortical and subcortical brain regions of near-term fetal sheep is modulated by vagus nerve manipulation during 54 h following the induction of a fetal inflammatory response with lipopolysaccharide (LPS). Vx+LPS represents combined Vx+LPS400 and Vx+LPS800 groups, since no differences regarding rCBF patterns were observed between the two subgroups. Y axis, Cerebral blood flow (ml/min/g). Sample sizes: Control, n=3; LPS, n=4; Vx+LPS, n=6; Vx+efferent VNS, n=5; Vx+afferent VNS, n=5.

**Discussion**

Complete cervical vagal denervation alters microglial phenotype in an LPS-dose-dependent manner which we refer to as hormetic due to a pronounced reversal of the microglial properties with a doubling of the LPS dose. This behavior is in line with the other reported systemic and organ-specific responses to LPS under conditions of vagotomy.[22] VNS further modulates microglial morphology in a brain-region-specific, i.e., locoregional, manner, as we demonstrate for the subregions of the hippocampus, a key vulnerable region involved in neurodegenerative processes. Vagus nerve manipulations also alter rCBF further highlighting the complex effects VNS exerts on brain function.

*Vagotomy reduces the activated state of microglia exposed to systemic inflammation induced by LPS*

Vagotomy combined with low-dose systemic LPS treatment appeared to reduce the 'activated' phenotype in microglia of the fetal sheep hippocampus. Increased Iba-1 expression has been shown to correlate with microglia's membrane ruffling to morphologically change from 'resting' to an activated amoeboid-shaped phenotype.[38] Here, we show that Vx+LPS400 treatment reduced Iba-1 mean fluorescent intensity for the whole microglia in the CA1 and for processes only in CA2/3 and DG, relative to control animals. Additionally, a decreased microglial soma to whole cell ratio was observed with Vx+LPS400 treated animals compared to controls in all 3 sub-regions. In the CA2 subregion, Vx+LPS400 treatment had this effect also compared to LPS only which underscores the impact of complete cervical vagal denervation on microglial morphology under conditions of neuroinflammation. In order to perform a variety of functional roles throughout development, prenatal microglia typically maintain an 'activated' amoeboid morphology with rounder, larger soma and short branching. This result suggests that vagotomy and low-dose LPS treatment favors the 'resting' microglia phenotype with smaller soma and more elaborate branching. While low-dose LPS exposure is sufficient to trigger functional microglial activation as presented in this model,[29] fetal microglia acquire a 'resting' phenotype in response to withdrawn cholinergic input from the vagus nerve. In a male adult rat model of subdiaphragmatic vagotomy, Gallaher et al. reported at 42 days following the vagotomy a mixed pattern of locoregional microglial activation (solitary tract (NTS), dorsal motor nucleus of the vagus nerve (DMV), and nodose ganglia (NG)) or silencing (spinal cord) measured by Iba-1 mean intensity.[39] Hofmann et al. identified the effects of unilateral right vagotomy after seven days in adult male rats on astrocytes and microglia populations in NTS.[40]

It is likely that there is a temporal profile to the locoregional activation pattern. This should be explored systematically in future studies: brain-regional, time-specific and developmental effects must be modeled in relation to vagus nerve activity and additional effects of systemic inflammation such as LPS exposure. Sex effects have been widely acknowledged for the inflammatory response patterns and will likely play a role in the physiology of vagal control of neuroinflammation as well.[41,42]

*Under the condition of vagotomy, Iba-1 expression in the fetal hippocampus depends on LPS dosage*

For fetal sheep receiving vagotomy, Iba-1 expression in the hippocampus is dependent on LPS dosage. We demonstrated that Iba-1 signal intensity was increased in Vx+LPS800 compared to Vx+LPS400 in whole microglia, soma, and processes from all 3 sub-regions. It is possible that a higher dose of LPS triggers sufficient systemic inflammation and cytokine production that disrupts the blood-brain barrier, therefore prompting a more reactive microglial phenotype. This hormetic effect under conditions of systemic inflammation with vagal withdrawal could impact neuroinflammatory response corresponding to the severity of chorioamnionitis.[22]

*Afferent or efferent VNS restores microglial phenotype*

Compared to Vx+LPS400 treatment alone, intermittent efferent and afferent VNS on top of the condition of Vx+LPS400 restores the levels of microglial activation to an extent similar to control and LPS.[29] When efferent VNS was added compared to Vx+LPS400 only, an increased microglia cell count in the CA1 was observed, suggesting that more microglia have either migrated to this region or proliferated to regulate immunological activity. Similarly, Vx+LPS400 animals that received efferent or afferent VNS showed an increased microglial soma:cell ratio. Although increased microglial soma size has been suggested to be a typical sign of microglial reactivity, there were no significant differences in soma size between these three groups. Therefore, the increased ratio suggests that microglia are undergoing a stage of the activation process where processes have been retracted.

Afferent and efferent VNS treatments lead to smaller soma size compared to treatments expected to cause inflammation: LPS, Vx+LPS800. That is compatible with the general notion that VNS is anti-inflammatory, especially in the hippocampus, here shown for the VNS effect on neuroinflammation.[43,44] It is important to note that the route of VNS had no measurable effect on the resulting microglial phenotype. This is important for the practical use of VNS where selective efferent or afferent stimulation can be more challenging to fully achieve, albeit progress has been made in this regard.[45]

These findings are contrasted by the pronounced systemic inflammatory response seen with afferent VNS compared to efferent VNS. Future work will show if other brain regions may show differences from the hippocampal behavior in response to LPS with afferent VNS. The rCBF patterns were more aligned with the observed systemic inflammatory responses differences. Future studies should explore these effects further.

*VNS modulates rCBF*

Vagus nerve manipulation results in changes of rCBF specific to cortical versus subcortical brain regions. This finding indicates that the vagus nerve exerts some control on the redistribution of rCBF, at least under conditions of unperturbed cerebral auto-regulation, as is likely the case in the presently described experiments: the blood pressure, especially the diastolic blood pressure did not drop to septic levels in these experiments as we reported elsewhere.[22]

Specifically, vagotomy followed by induction of fetal inflammatory response results in an amplified redistribution of rCBF in favor of subcortical perfusion. In contrast, efferent VNS reduces the rCBF to below control levels for the cortical matter, while afferent VNS diminishes the rise of rCBF in the subcortical, but not cortical matter.

Future research will attempt to causally relate the effects of the vagal activity on locoregional patterns of cerebral perfusion and neuroinflammation and how these patterns correlate to acute neurological and neurodevelopmental outcomes. The ability to modulate vagal activity via VNS provides here a novel therapeutic modality for early intervention to reduce sequlae from perinatal brain injury.

*Therapeutic implications for VNS in the treatment of neurodegenerative disorders*

Going beyond perinatal development, the interplay between the CAP and microglial cells offers enormous therapeutic potential. Early discoveries involving both preclinical and clinical data have implicated the CAP in neurodegenerative diseases like Parkinson's, explained in part by interactions with microglial cells. The Braak hypothesis for Parkinson's pathogenesis has demonstrated that α-synuclein originating in the Meissner's and Auerbach's plexuses of the gastrointestinal system is transported in a retrograde fashion up the vagus nerve to brain regions like the substantia nigra pars compacta in the midbrain, the primary site of dopaminergic neurons' destruction in Parkinson's.[46] In neural tissue, misfolded α-synuclein contributes to a chronic neuroinflammatory response. Neuropathological post-mortem analysis has demonstrated increased levels of pro-inflammatory factors, microglia activation, and T-cell activation. The pro-inflammatory state, driven in part by α-synuclein, is believed to be both directly cytotoxic to dopaminergic neurons and catalytic for neurodegeneration. Inflammatory signaling increases blood-brain barrier permeability, which allows for increased α-synuclein transport into the brain. The neuroinflammatory response within the brain then drives α-synuclein aggregation, further perpetuating activation of the innate and adaptive immune responses and repeating the cycle. Thus, the neuroinflammatory response seen in Parkinson's disease, which is largely driven by microglial cells, essentially acts as a closed-loop system for neurodegeneration.[47] Animal modeling and retrospective cohort analysis in humans have substantiated the Braak hypothesis, demonstrating that vagotomy is associated with decreased incidence of Parkinson's disease.[48] As suggested by the previous modeling of the Braak hypothesis, the likely mechanism for this finding is that vagotomy prevents α-synuclein transport to the brain; therein limiting the neuroinflammatory response that drives neurodegeneration.[47]

Microglial cells are the immune cells primarily responsible for neuronal cell death in Parkinson's disease. As outlined above, evidence suggests in part that a-synuclein activates microglial cells triggering the CNS neuroinflammatory response.[47] Interestingly, α-synuclein structure corresponds with the functional role of microglial cells in this neuroinflammatory response, and thus disease progression. The more classical mutant structures of α-synuclein, like A30P or A53T, lead to a proinflammatory activation state with subsequent neurodegeneration. However, monomeric a-synuclein results in activated microglial cells that serve a neuroprotective role.[47] The functional diversity demonstrated through the Braak mechanism for Parkinson's illustrates why there is hope to one day clinically manipulate the relationship between the CAP and microglial cells to counter neurodegeneration via VNS.

The excitement surrounding the functional diversity of microglial cells and the potential therapeutic implications it offers are not limited to neurodegeneration seen in Parkinson's disease. Microglial cells have been implicated across multiple disease states. One proposed mechanism for neurodegeneration in other pathologies is the re-activation of complement-mediated developmental synapse pruning by microglial cells. During normal neural development, microglia rely on C1q tagging of synapses to identify phagocytic targets to allow trimming of the neuronal network. As the fetus approaches term, such activity is down-regulated.[49] However, mice models for pathologies like Alzheimer's, Pick's disease (frontal-temporal dementia), acute West Nile encephalopathy, and aging have all demonstrated aberrant interactions between the classical complement system and activated microglial cells leading to synaptic destruction. The functional impact of unregulated synaptic destruction by microglia presents as the cognitive decline that hallmarks these disease processes.[15]

The growing body of evidence makes clear that there is still much to be learned in regard to the functional diversity of microglial cells and their role in many CNS pathologies. Data indicate that microglial cells have the capacity to serve a multitude of roles, and which role they take on can be guided by cues from their microenvironment. Additionally, the vagus nerve is more involved in the regulation of the innate immune system than previously thought. In fact, it may provide a mechanism for the modulation of microglial cell microenvironments; a potential pathway toward a clinical intervention for neurodegeneration and other conditions where neuroinflammation is a factor.

*Implications of VNS in studies of perinatal brain development and function: new insights and opportunities*

There remain many open questions. On the systems level, we need to understand the relationship between the vagal activity and microglial phenotype. The expression of α7nAChRs on microglia appears to be dynamic in relation to the vagal activity.[2] How does this dynamics drive the microglial phenotype and can we leverage VNS to modulate microglial properties for salutory purposes? Vagal activity also appears to modulate the systemic metabolic state, at least as far as we can observe under conditions of mild to moderate inflammation as may occur during chorioamnionitis.[22]

On the cellular level, the in vitro α7nAChR stimulation modulates the immunometabolic phenotypes of microglia and astrocytes in the perinatal brain.[9,29,50]

Together, these findings have led to the neuro-immunometabolic hypothesis of the etiology of autism spectrum disorder (ASD), a condition whose origins have been associated with intrauterine adversities such as neuroinflammation and chronic stress.[51]
We discuss the implications for the developmental origins of neurodegenerative disorders in a separate subsection.

Methodologically, it is important to emphasize that we propose an approach whereby not only mean intensity levels of Iba-1 positive microglia are quantified in relation to an intervention, but also their complex soma-processes' morphology is accounted for. We suggest that a particular attention be paid in future work to the cell morphometric assessments of microglial plasticity, especially in relation to VNS. That is also likely true for understanding the VNS effects on the plasticity of other glial cell populations, notably the astrocytes.

VNS is usually thought of as an *exogenous* process, but we also observed a dynamic relationship between microglia and CAP activity under *endogenous* "VNS" conditions. These are the types of VNS that arise from physiological or pathophysiological states such as fetal acidemia or inflammation which directly stimulate the vagus nerve.[2]

VNS can therefore be thought of as being exogenous or endogenous in origin, with its own set of effects on systems and cellular scales of physiological organization.

This view raises the question how the exogenous or endogenous VNS effects can be decoded indirectly from readily available signal sources such as electrocardiogram (ECG)-derived heart rate variability (HRV), a notion we refer to as the HRV code.[52] In the related contribution, on the systems level, we present a direct (vagus electroneurogram, VENG) approach to decoding vagus nerve activity and demonstrate its relationship to the indirect, HRV-based approach.[53,54] The ECG-derived HRV inflammatory index tracks the rise of the systemic inflammatory response and the changes in the VENG using the same underlying machine learning algorithm.

An important limitation of the presented work and inferences is the lack of vagotomy-only group as a sham to Vx+LPS groups in the presented experiments. That remains subject of future work.

*Significance*

In summary, VNS modulates microglial phenotype in hippocampus of the developing brain. This may be leveraged for acute modulation of perinatal brain injury.[55] Moreover, vagus nerve activity appears to have a long-term impact on the neurodevelopment and lifetime risk for neurodegenerative disorders. Future research needs to seek for optimal VNS treatment regimes for various adversities to ensure that these do not alter the long-term microglial phenotype unfavourably.

**References**

1. Pavlov, V. A. & Tracey, K. J. Bioelectronic medicine: Preclinical insights and clinical advances. *Neuron* (2022) doi:10.1016/j.neuron.2022.09.003.

2. Frasch, M. G. *et al.* Decreased neuroinflammation correlates to higher vagus nerve activity fluctuations in near-term ovine fetuses: a case for the afferent cholinergic anti-inflammatory pathway? *J. Neuroinflammation* **13**, 103 (2016).

3. Pavlov, V. A. *et al.* Central muscarinic cholinergic regulation of the systemic inflammatory response during endotoxemia. *Proc. Natl. Acad. Sci. U. S. A.* **103**, 5219–5223 (2006).

4. Novellino, F. *et al.* Innate Immunity: A Common Denominator between Neurodegenerative and Neuropsychiatric Diseases. *Int. J. Mol. Sci.* **21**, (2020).

5. Carnevale, D. Neuroimmune axis of cardiovascular control: mechanisms and therapeutic implications. *Nat. Rev. Cardiol.* **19**, 379–394 (2022).

6. Jakob, M. O., Murugan, S. & Klose, C. S. N. Neuro-Immune Circuits Regulate Immune Responses in Tissues and Organ Homeostasis. *Front. Immunol.* **11**, 308 (2020).

7. Diamond, B. & Tracey, K. J. Mapping the immunological homunculus. *Proc. Natl. Acad. Sci. U. S. A.* **108**, 3461–3462 (2011).

8. Rosas-Ballina, M. *et al.* Acetylcholine-synthesizing T cells relay neural signals in a vagus nerve circuit. *Science* **334**, 98–101 (2011).

9. Cortes, M. *et al.* α7 nicotinic acetylcholine receptor signaling modulates the inflammatory phenotype of fetal brain microglia: first evidence of interference by iron homeostasis. *Sci. Rep.* **7**, 10645 (2017).

10. Frasch, M. G. & Nygard, K. L. Location, Location, Location: Appraising the Pleiotropic Function of HMGB1 in Fetal Brain. *J. Neuropathol. Exp. Neurol.* **76**, 332–334 (2017).